\documentclass[aps,prd,nofootinbib]{revtex4}
\usepackage{graphicx}
\usepackage{amsfonts}
\usepackage{amsmath}
\usepackage{amssymb}

\begin{document}

\title{On a Class of Harko--Kovacs--Lobo Wormholes}
\author{R.Kh. Karimov}
\email{karimov_ramis_92@mail.ru}
\affiliation{Zel'dovich International Center for Astrophysics, Bashkir State Pedagogical University, 3A, October Revolution Street, Ufa 450008, RB, Russia}
\author{R.N. Izmailov}
\email{izmailov.ramil@gmail.com}
\affiliation{Zel'dovich International Center for Astrophysics, Bashkir State Pedagogical University, 3A, October Revolution Street, Ufa 450008, RB, Russia}
\affiliation{Institute of Molecule and Crystal Physics, Ufa Federal Research Centre, Russian Academy of Sciences, Prospekt Oktyabrya 151,  450075 Ufa, RB, Russia}
\author{K.K. Nandi}
\email{kamalnandi1952@rediffmail.com}
\affiliation{Zel'dovich International Center for Astrophysics, Bashkir State Pedagogical University, 3A, October Revolution Street, Ufa 450008, RB, Russia}

\date{18 October 2022}

\begin{abstract}
The Harko, Kov\'{a}cs, and Lobo wormhole (HKLWH) metric contains two free parameters: one is the wormhole throat $r_{0}$, and the other is a dimensionless deviation parameter $\gamma$ with values $0<\gamma <1$, the latter ensuring the needed violation of the null energy condition at the throat. In this paper, we study the energetics of the HKLWH and the influence of $\gamma$ on the tidal forces in the Lorentz-boosted frame. Finally, we apply\ a new concept, namely, the probabilistic identity of the object observed by different external observers in terms of the Fresnel coefficients derived by Tangherlini. The intriguing result is that observations can differ depending on the location of the observer, i.e., there is a nonzero probability that the HKLWH will be identified as a black hole even when $\gamma \neq 0$.
\end{abstract}

\maketitle


\section{Introduction}
\label{sec:intro}
Wormholes (WHs) were conceived as a particle model by Einstein and Rosen (the well known Einstein--Rosen bridge) in 1935  \cite{Einstein:1935}. Ellis  \cite{Ellis:1973} and  Bronnikov  \cite{Bronnikov:1973}, independently in 1973, found WH solutions threaded by an Einstein minimally coupled scalar field with a negative kinetic term in the Lagrangian. After the seminal work by Morris and Thorne in 1988  \cite{Morris:1988}, there was a revival of interest in the subject among the physics community. WHs are topological handles connecting two spacetimes that have not yet been ruled out by experiments. In fact, in the context of semiclassical quantum gravity, the minimal two-surface is not the horizon of a black hole (BH) but the throat of a WH  \cite{Barthiere:2018}. The possibility of the appearance of Planck-sized WHs in quantum gravity is well discussed in the literature \cite{Nandi:2004, Roman:1986}. Moreover, there has been much work on the possibility of observationally distinguishing classical WHs from BHs by means of various diagnostics, such as an accretion disc, gravitational lensing, etc.; see, for example,  \cite{Karimov:2019, Harko:2009, Tsukamoto:2017, Shaikh:2017, Shaikh:2018, Tsukamoto:2020a, Tsukamoto:2020b, Korolev:2020, Yusupova:2021, Stuchlik:2021, Paul:2021, Radinschi:2021, Alencar:2021, Bronnikov:2021, Fabris:2021, Jusufi:2022, Sokoliuk:2022}.

The purpose of the present paper is to study the features of the HKLWH in reasonable detail, emphasizing the role of $\gamma$ in various physical effects that include energetics, tidal effects, and most interestingly, the probabilistic identification of the type of the object as an outcome of probes by photons, i.e., we calculate the probability of the HKLWH being identified as a WH or a BH by different observers using probability flows of photons as probes.

To achieve this, we adopt Tangherlini's  \cite{Tangherlini:1975} idea of the probabilistic scattering of photons, which is as follows. Classically, a photon motion is a regular deterministic wave propagation into an optical medium with reflection and transmission amplitudes; however, Tangherlini introduced \textit{an indeterministic condition instead of a deterministic one at the interface}. The indeterminacy was introduced by assuming a hypothetical arrangement in which individual photons are incident on widely separated media at random intervals of time. As a consequence, the physical flow of energy into the medium cannot be associated with a physical flow of energy but some kind of {probability flow} associated with each photon. In other words, the Fresnel coefficients are determined in terms of a statistical ensemble average over a large number of replica media having the same index $n$---with one photon for each replica. Because of the assumed independence of collisions at the medium interface, the coefficients are the same for one photon as for $N$ photons  \cite{Tangherlini:1975}.

The above ingenious idea of Tangherlini does not use Planck's constant but an experimentally verified photon momentum increase in the refractive medium, namely, $p^{\prime} = np$, $n\geq 1$. Equivalently, the de Broglie relation $p^{\prime}\lambda ^{\prime} = p\lambda =$ constant (\textit{without Planck's constant }$\hslash$), together with the well-known reduction in the wavelength $\lambda^{\prime} = \lambda/n$, then leads to $p^{\prime }=np$, which was used as a starting point by Tangherlini  \cite{Tangherlini:1975} to derive the Fresnel coefficients $R$ and $T$. Remarkably, the same Fresnel coefficients can also be deduced  from the standard quantum mechanical treatment of the Schr\"{o}dinger equation for a certain potential well, which reinforces the validity of the assumed probability flow in Tangherlini's scheme. Loosely speaking, we can say that the scheme yields ``quantum mechanics without quantum mechanics''. That is why Tangherlinini called his approach a ``pre-quantal statistical formulation''. No information is reduced, and the new idea is that the photon motion into the medium is raised from the classical deterministic level to a so-called pre-quantal statistical flow of energy without Planck's constant. For further explanation,  see the original \mbox{paper  \cite{Tangherlini:1975}}. The purpose of Tangherlini's approach  \cite{Tangherlini:1975}, as enumerated by the author himself, is that ``The approach should be of value in further clarifying questions about the foundations of quantum mechanics as well as fundamental questions about the interaction of light \mbox{and matter''.}

The expected outcome in the present study is that an external observer has the probability to observe the probed centrally gravitating object either as a BH or a WH, thus forming a two-point sample space. The indeterministic effect of Tangherlini's approach is dictated {only} by the WH geometry, which subsumes the source energy distribution in the spirit of general relativity, no matter whether the source satisfies the energy conditions or not. A photon is propagating (in the statistical sense, as explained above) through the exterior energy-violating region of the HKLWH, portrayed as an ``effective optical medium''. We make a similar hypothetical arrangement to that explained above but now involving a gravitational ``effective optical medium'' to probe the nature of the central object. If every photon is reflected back by the probed surface (implying $R$ $=1$, $T=0$), the observer sending the pulse would identify the surface to be a BH horizon with {certainty}. The ground for taking $R$ $=1$, $T=0$ as a hallmark of a BH is that these values follow directly from the conventional quantum mechanical treatment using the Schr\"{o}dinger equation for a certain potential well  \cite{Bohm:1951}. On the other hand, if all of the photons transmit through the surface, so that $R$ $=0$, $T=1$, then the observer identifies the surface as a WH throat with certainty. If some photons are reflected and some transmitted, then depending on the value of $\gamma$ and location of the observer, the latter is {more likely} to identify the object as a BH (if $R>T)$ or a WH (if $R<T$) in a large number of probes. Thus, for any $\gamma \neq 0$, however tiny, the ($R,T$) values would differ leading correspondingly to the different outcomes in the two-point sample space (BH or WH). The essence of all this is that $\gamma \neq 0$ does not necessarily lead to the observation of a HKLWH but, statistically speaking, can also lead to a BH.

To apply Tangherlini's idea in gravity, we shall first portray the HKLWH spacetime as a gravitational ``effective optical medium'' with index $n(r)$ as sensed by photons sent by asymptotic and another index $\widetilde{n}(r)$ sensed by photons sent by near-throat observers\footnote{%
Yet another different index $N(r)$ is sensed by a massive particle in motion  \cite{Evans:2001, Evans:1996a, Evans:1996b, Alsing:2001, Alsing:1998, Evans:1986}. In the present paper, we deal only with  photon motion taking into account that, in general relativity, the observations depend on the location of the observer. The photon motion in the effective medium also gives rise to the interesting possibility of the gravitational analogue of the Fizeau effect in the ``effective optical medium''  \cite{Nandi:2003}.} \cite{Nandi:2016}. We emphasize that Tangherlini's formulation was originally developed for an ordinary optical medium, but its present application to the gravitational effective optical medium is justified on the famous Pound--Rebka experiment in a gravity field. The reason for this assertion can be seen from the connection that, while the required momentum increase law $p^{\prime }=np$, which is already an experimentally confirmed fact in an ordinary optical medium  \cite{Jones:1951, Jones:1954, Ashkin:1973}, the Pound--Rebka experiment in gravity can also be nicely reinterpreted as yet another momentum increase law but in the effective optical medium with index $n(r)$, as shown in detail in  \cite{Nandi:2016}.

The paper is organized as follows: In Section \ref{s2}, we study the energetics of the HKLWH, and in Section \ref{s3}, we study the influence of $\gamma$ on the tidal forces in the Lorentz-boosted frame. Finally, in Section \ref{s4}, we apply a new concept, namely, the identification of the HKLWH by different observers in terms of the probabilistic Fresnel coefficients. We shall use a gravitational refractive medium representing the HKLWH to calculate the influence of $\gamma$ on these coefficients. Section \ref{s5} concludes the paper. We take units $G=c=1$, unless specifically restored.

\section{HKLWH and Its Energetics}\label{s2}
The Morris and Thorne \cite{Morris:1988} WH has a generic form in standard coordinates
\begin{equation}
d\tau ^{2}=e^{2\Phi (r)}dt^{2}-\frac{dr^{2}}{1-b(r)/r}-r^{2}d\Omega ^{2},
\end{equation}%
where $d\Omega^{2} \equiv d\theta^{2} + \sin^{2}{\theta}d\varphi^{2}$ represents the metric on a unit sphere. The function $\Phi (r)$ is called the redshift function,  $b(r)$ is  the shape function, and the root of $b(r_{0})=r_{0}$ is the throat radius, such that $r\in \lbrack r_{0},\infty)$. The material threading the WH has the density $\rho_{d}$ and pressures ($p_{r}$, $p_{t}$) (components of transverse pressures, $p_{\theta}$, $p_{\varphi}$ are collectively called $p_{t}$) given by
\begin{eqnarray}
\rho _{d} &=& \frac{1}{8\pi}\frac{b^{\prime }}{r^{2}}, \\
p_{r} &=& \frac{1}{8\pi} \left[2 \left(1-\frac{b}{r}\right) \frac{\Phi^{\prime}}{r} - \frac{b}{r^{3}}\right], \\
p_{t} &=& \frac{1}{8\pi }\left( 1-\frac{b}{r}\right) \left[ \Phi ^{\prime
\prime }+\left( \Phi ^{\prime }\right) ^{2}-\frac{b^{\prime }r-r}{%
2r^{2}\left( 1-b/r\right) }\Phi ^{\prime }-\frac{b^{\prime }r-r}{%
2r^{3}\left( 1-b/r\right) }+\frac{\Phi ^{\prime }}{r}\right].
\end{eqnarray}

Harko, Kov\'{a}cs, and Lobo  \cite{Harko:2008} considered a wormhole metric supported by exotic matter. It is well known that exotic matter is a type of matter that violates one or more energy conditions, especially the Null Energy Condition (NEC)  \cite{Lobo:2017, Nandi:2004b}. The violation of the NEC is the minimal condition for the existence of traversable WHs  \cite{Lobo:2017}. The Harko--Kov\'{a}cs--Lobo wormhole (HKLWH) metric  \cite{Harko:2008} is given in standard coordinates as
\begin{equation}
ds^{2} = e^{-2r_{0}/r}dt^{2} - \frac{dr^{2}}{1-r_{0}[1+\gamma (1-r_{0}/r)]/r} - r^{2}d\Omega^{2},
\end{equation}%
where $0<\gamma <1$ is a constant, and $r_{0}$ is the minimum radius defining the wormhole throat. The metric (5) describes a wormhole geometry with two identical asymptotically flat regions joined together at the throat $r_{0}>0$. The redshift function of the wormhole is $\Phi (r)=-r_{0}/r$, and the shape function is $b(r)=r_{0}+\gamma r_{0}(1-r_{0}/r)$. In  \cite{Harko:2008}, it was shown that particles moving in circular orbits around wormholes are stable, due to the outward gravitational repulsion.

To avoid the presence of event horizons to ensure communications between the two connecting universes, $\Phi (r)$ is constrained to be finite throughout the coordinate range. At the throat $r_{0}$, one has $b(r_{0})=r_{0}$, and a fundamental property is the flaring-out condition given by $(b^{\prime 2}<0$, which is provided by the geometric method of \mbox{embedding  \cite{Morris:1988}}. Most wormholes are made up of exotic matter that violates one or more energetic \mbox{conditions  \cite{Lobo:2017, Harko:2013, Parsaei:2019}}. Here, we will check the null (NEC), strong (SEC), and dominant (DEC) energy conditions following Rodrigues et al.  \cite{Rodrigues:2016}
\begin{eqnarray}
&&NEC_{1}(r) = \rho_{d} +  p_{r}\geq 0, \quad NEC_{2}(r) = \rho_{d} +  p_{t}\geq 0, \\
&&SEC(r) = \rho_{d} + p_{r} + 2p_{t}\geq 0, \\
&&WEC_{1,2}(r) = \rho_{d} + p_{r}\geq 0, \quad WEC_{2}(r) = \rho_{d} + p_{t}\geq 0, \\
&&DEC_{1}(r) = \rho_{d}\geq 0, \\
&&DEC_{2}(r) = \rho_{d} - p_{r}\geq 0, \quad DEC_{3}(r) = \rho_{d} - p_{t}\geq 0.
\end{eqnarray}

After putting (5) into (6)--(10), we obtain
\begin{eqnarray}
&&NEC_{1}(r)=WEC_{1}(r)=-\frac{r_{0}}{r^{3}}\left[ 1-\gamma -2(1+\gamma ) \frac{r_{0}}{r}+2\gamma \frac{r_{0}^{2}}{r^{2}}\right] , \\
&&NEC_{2}(r)=WEC_{2}(r)=-\frac{r_{0}}{2r^{3}}\left[ 1-\gamma -(5+3\gamma ) \frac{r_{0}}{r}+2(1+3\gamma )\frac{r_{0}^{2}}{r^{2}}-2\gamma \frac{r_{0}^{3}}{r^{3}}\right] , \\
&&SEC(r)=-\frac{r_{0}}{r^{3}}\left[ 2(1-\gamma )-(7+3\gamma )\frac{r_{0}}{r} + 2 (1+4\gamma )\frac{r_{0}^{2}}{r^{2}}-2\gamma \frac{r_{0}^{3}}{r^{3}}\right], \\
&&DEC_{1}\left( r\right) =\frac{\gamma r_{0}^{2}}{8\pi r^{4}}, \\
&&DEC_{2}\left( r\right) =\frac{r_{0}}{r^{3}}\left[ 1-\gamma -2\frac{r_{0}}{r}+2\gamma \frac{r_{0}^{2}}{r^{2}}\right] , \\
&&DEC_{3}\left( r\right) =\frac{r_{0}}{2r^{3}}\left[ 1-\gamma -(5-\gamma) \frac{r_{0}}{r}+2(1+3\gamma )\frac{r_{0}^{2}}{r^{2}}-2\gamma \frac{r_{0}^{3}}{r^{3}}\right] .
\end{eqnarray}

\begin{figure}
\centering
\includegraphics[width=18.0cm]{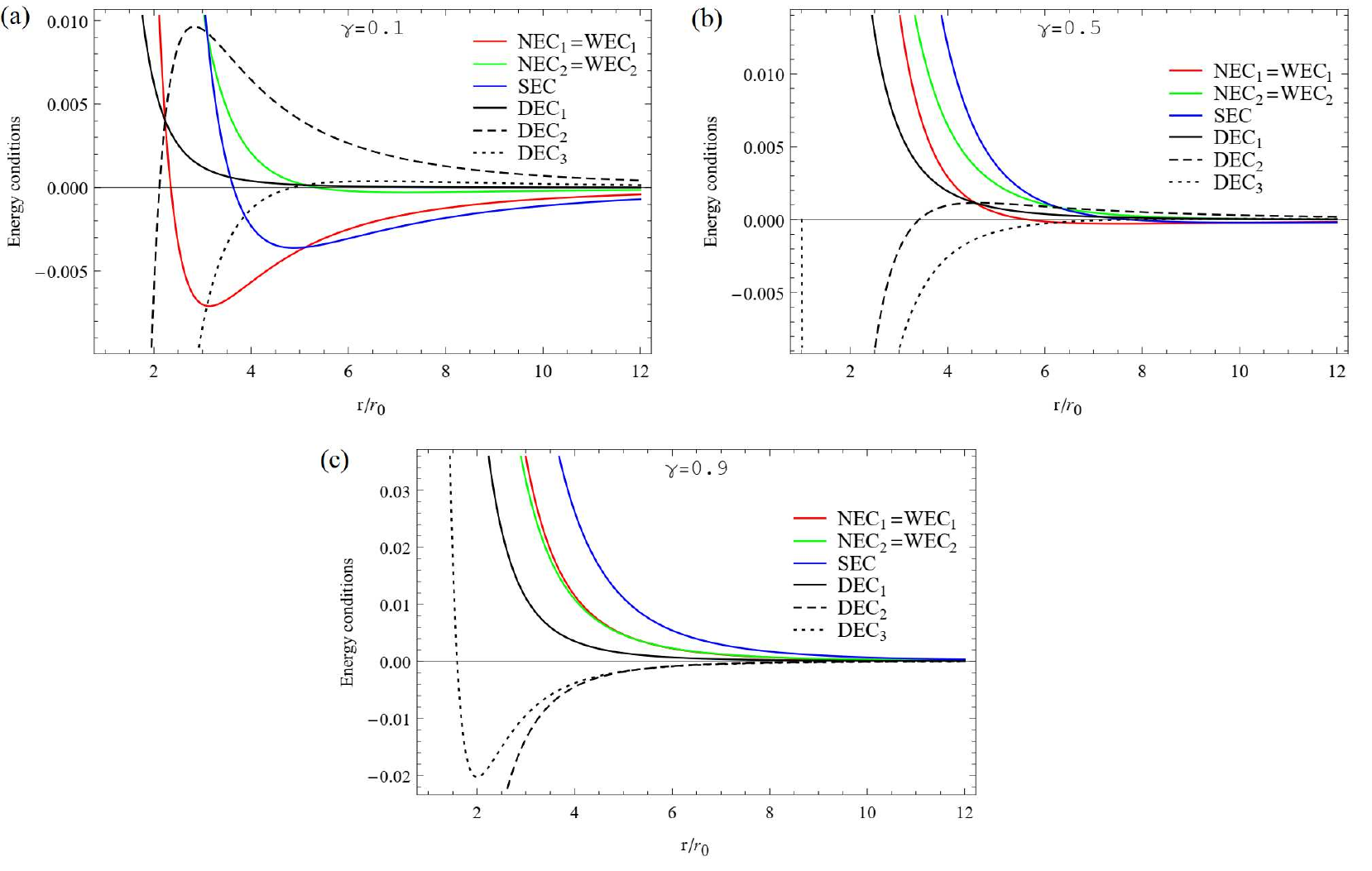}
\caption{The energy conditions of the HKLWH as $\gamma = 0.1$ (\textbf{a}, top left hand), $\gamma = 0.5$ (\textbf{b}, top right hand), and $\gamma = 0.9$ (\textbf{c}, bottom).\label{fig1}}
\end{figure}

Figure \ref{fig1} shows the dependence of the energy conditions of the HKLWH on the radial coordinate $r$ for different values of the dimensionless metric parameter $\gamma $. $NEC$ is violated near the throat of the wormhole. In this case, the range of the $NEC$ violation directly depends on the value of $\gamma $. The reverse picture is observed for $NEC_{2}$ and $DEC_{2}$; as $\gamma$ increases, the range of violation of energy conditions decreases. The energy conditions $SEC$ and $DEC_{2}$ are not violated at any $\gamma$ values. An interesting case is observed in the case of $DEC_{3}$: at $\gamma \leq 0.5$, the energy condition is violated near the throat radius; at $\gamma > 0.5$, a gap appears near the throat, where the energy condition is not violated.

From the integration of the $00$-component of the Einstein field equations (see,  \cite{Lynden-Bell:2007a}), we obtain%
\begin{equation}
\frac{\partial m}{\partial r}=4\pi r^{2}T_{0}^{0}=4\pi r^{2}\rho _{d}
\end{equation}%
for a metric ansatz%
\begin{equation}
d\tau ^{2}=\xi (r,t)dt^{2}-\frac{dr^{2}}{1-2m(r,t)/r}-\left( r^{2}d\theta^{2}+\sin ^{2}{\theta }d\varphi ^{2}\right),
\end{equation}%
where $m(r,t)$ is the mass function, and $T_{0}^{0}(r,t)$ is the energy density of matter in the rest frame $\left( r,\theta ,\varphi \right) $constant. We see the genesis of the volume measure $4\pi r^{2}$ in Equation (17). With the volume measure $4\pi r^{2}dr$, the mass function in the static case is given by
\begin{equation}
m=\frac{1}{8\pi }\int_{r_{0}}^{\infty }4\pi \rho _{d}r^{2}dr=\frac{\gamma
r_{0}}{2},
\end{equation}%
which is the ADM mass. If one introduces the radial fluid motion with a velocity $v$ of the source matter, then the static observers in the rest frame $\left(r, \theta, \varphi \right) $ constant see a Lorentz transformed $T_{00} = \frac{\rho _{d}+p_{r}v^{2}}{1-v^{2}}$. The Lorentz contraction due to fluid motion gives rise to a mechanical energy flux within $r=$ const passing through the hypersurface $t=$ const. However, even for a static configuration (such that $v=0$), Lynden-Bell et al.  \cite{Lynden-Bell:2007a} asserted that the mechanical energy $E_{M}$ in the spacetime, whatever constitutes the $T_{\mu \nu }$, is still given by
\begin{equation}
E_{M}=\frac{1}{8\pi}\int_{r_{0}}^{\infty }4\pi \rho_{d}(g_{rr})^{1/2}r^{2}dr.
\end{equation}

Lynden-Bell et al.\cite{Lynden-Bell:2007a} made a further assertion that ``\textit{A comparison between \linebreak Equations  (19) and (20) illustrates that (19) is seductively like the classical relationship between density and mass, but in fact conceals all the complications beneath a cloak of apparent clarity}.'' For instance, $4\pi r^{2}$ is {not} the proper volume measure, while the correct one is $4\pi (g_{rr})^{1/2}r^{2}dr=dV$.

The ADM mass $m$ is the net mass-energy that contains the self-gravitational energy $E_{G}$, together with the other mechanical energies, such as the rest mass-energy and the internal energy collectively denoted by $E_{M}.$ Thus, Lynden-Bell et al.  \cite{Lynden-Bell:2007a} advocated that $E_{G}=m-E_{M}<0$ should be the generic definition of self-gravitational energy that can be explicitly computed for any given system.

In the present case, we find that
\begin{eqnarray}
E_{M} &=&\frac{\sqrt{\gamma }r_{0}}{2}\log \left[ \frac{1+\sqrt{\gamma }}{1-%
\sqrt{\gamma }}\right] >0\quad \Rightarrow \quad \\
E_{G} &=&\frac{\gamma r_{0}}{2}\left\{ 1-\frac{1}{\sqrt{\gamma }}\log \left[
\frac{1+\sqrt{\gamma }}{1-\sqrt{\gamma }}\right] \right\} <0.
\end{eqnarray}%

This shows that one side of the mouth with an asymptotic mass $m(r\rightarrow \infty ) = \frac{\gamma r_{0}}{2}$ is attractive. By extending the coordinate path to cover the entire real line, $-\infty <\overline{r}<+\infty $, exposing the two asymptotically flat regions, it can be shown that the negative mass mouth is repulsive, i.e., it has $E_{G}>0$ . Thus, while one side attracts the traveller in, the other side repels the traveller out, as it should be.

\section{Influence of \boldmath$\gamma$ on the Tidal Forces in the Lorentz-Boosted Frame}\label{s3}
We start with the general form of a static spherically symmetric physical metric:
\begin{equation}
d\tau ^{2}=\frac{F(r)}{G(r)}dt^{2}-\frac{dr^{2}}{F(r)}-R^{2}(r)[d\theta^{2}+\sin ^{2}\theta d\varphi ^{2}].
\end{equation}%

For a traveler in a static orthonormal frame, we shall denote the only nonvanishing components of the Riemann curvature tensor as $R_{0101}$, $R_{0202}$, $R_{0303}$, $R_{1212}$, $R_{1313}$, and $R_{2323}$. Radially freely falling observers with conserved energy $E$ are connected to the static orthonormal frame by a local Lorentz boost with an instantaneous velocity given by
\begin{equation}
v=\left[ 1-\frac{F}{GE^{2}}\right] ^{1/2}.
\end{equation}%

Then, the nonvanishing Riemann curvature components in the Lorentz-boosted frame with velocity $v$ are ($k=2,3$):
\begin{eqnarray}
R_{\widehat{0}\widehat{1}\widehat{0}\widehat{1}} &=&R_{0101}, \\
R_{\widehat{0}\widehat{k}\widehat{0}\widehat{k}} &=&R_{0k0k}+\left(
R_{0k0k}+R_{1k1k}\right) \sinh ^{2}{\alpha }, \\
R_{\widehat{0}\widehat{k}\widehat{1}\widehat{k}} &=&\left(
R_{0k0k}+R_{1k1k}\right) \sinh {\alpha }\cosh {\alpha }, \\
R_{\widehat{1}\widehat{k}\widehat{1}\widehat{k}} &=&R_{1k1k}+\left(
R_{0k0k}+R_{1k1k}\right) \sinh ^{2}{\alpha },
\end{eqnarray}%
where $\sinh {\alpha }=v/\sqrt{1-v^{2}}$. The relative tidal acceleration $\Delta a_{\widehat{j}}$ between two parts of the traveler's body in the orthonormal basis is given by
\begin{equation}
\Delta a_{\widehat{j}}=-R_{\widehat{0}\widehat{j}\widehat{0}\widehat{p}}\xi^{\widehat{p}},
\end{equation}%
where $\overrightarrow{\xi }$ is the vector separation between the two parts  \cite{Nandi:2001b}. Thus the curvature components contributing to the tidal force on the traveler in the Lorentz-boosted frame are $R_{\widehat{0}\widehat{1}\widehat{0}\widehat{1}}$, $R_{\widehat{0}\widehat{2}\widehat{0}\widehat{2}}$, and $R_{\widehat{0}\widehat{3}\widehat{0}\widehat{3}}$. (Components in the coordinate basis are not required here).

The component $R_{\widehat{0}\widehat{2}\widehat{0}\widehat{2}}$ is given by
\begin{eqnarray}
R_{\widehat{0}\widehat{2}\widehat{0}\widehat{2}} &=&-\frac{1}{R}\left[ \frac{%
R^{\prime }}{2}\left( E_{s}^{2}G^{\prime }-F^{\prime }\right) \right] -\frac{%
1}{R}\left( R^{\prime \prime }G+\frac{R^{\prime }G^{\prime }}{2}\right)
E_{ex}^{2} \\
&=&R_{\widehat{0}\widehat{2}\widehat{0}\widehat{2}}^{(s)}+R_{\widehat{0}%
\widehat{2}\widehat{0}\widehat{2}}^{(ex)},
\end{eqnarray}%
where
\begin{equation}
E^{2}=\frac{F}{G}+\frac{F}{G}\left( \frac{\mathbf{v}^{2}}{1-\mathbf{v}^{2}}%
\right) =E_{\text{s}}^{2}+E_{\text{ex}}^{2},  \label{3.5}
\end{equation}%
where $E_{\text{s}}^{2}$ represents the value of $E^{2}$ in the static frame, and $E_{\text{ex}}^{2}$ represents the enhancement in $E_{\text{s}}^{2}$ due to the geodesic motion.

It is easy to verify that the term $|R_{\widehat{0}\widehat{2}\widehat{0}\widehat{2}}^{(s)}|$ actually represents the curvature component in the static frame, namely, $R_{\widehat{0}\widehat{2}\widehat{0}\widehat{2}}^{(s)}=R_{0202}$. Thus, only the term $R_{\widehat{0}\widehat{2}\widehat{0}\widehat{2}}^{(ex)}$ $[=\sinh ^{6}\alpha (R_{0202}+R_{1212})]$ represents the overall enhancement in the curvature in the Lorentz-boosted frame over the static frame. It is this part that needs to be particularly examined as the observer approaches the horizon. Note also that the energy $E^{2}$ is finite (it can be normalized to unity) and so are $E_{s}^{2}$ and $E_{ex}^{2}$. As the horizon is approached, $(F/G)\rightarrow 0$, $v\rightarrow 1$ such that $E^{2}\rightarrow E_{ex}^{2}$.

From Equations (5) and (23), we find
\begin{eqnarray}
F(r) &=&1-\frac{r_{0}}{r}\left\{ 1+\gamma \left( 1-\frac{r_{0}}{r}\right)
\right\} , \\
G(r) &=&\left[ 1-\frac{r_{0}}{r}\left\{ 1+\gamma \left( 1-\frac{r_{0}}{r}%
\right) \right\} \right] e^{\frac{2r_{0}}{r}}, \\
R(r) &=&r.
\end{eqnarray}

Using Equations (30), (33)--(35), we find%
\begin{equation*}
R_{\widehat{0}\widehat{1}\widehat{0}\widehat{1}}=-\frac{r_{0}^{2}}{2r^{5}}%
\left[ r(1+\gamma )-2\gamma r_{0}\right]
\end{equation*}%
\begin{equation}
R_{\widehat{0}\widehat{2}\widehat{0}\widehat{2}}^{(s)}=\frac{r_{0}}{r^{3}}%
\left( 1-\frac{r_{0}}{r}\right) \left( 1-\frac{\gamma r_{0}}{r}\right) ,
\end{equation}%
and
\begin{equation}
R_{\widehat{0}\widehat{2}\widehat{0}\widehat{2}}^{(ex)}=\frac{r_{0}}{2r^{3}}%
\left( 1-\gamma -2\frac{r_{0}}{r}+2\gamma \frac{r_{0}^{2}}{r^{2}}\right)
\left( \frac{v^{2}}{1-v^{2}}\right) ,
\end{equation}%
which, at the throat, $r=r_{0}$, yields
\begin{equation}
R_{\widehat{0}\widehat{1}\widehat{0}\widehat{1}}=\frac{\gamma -1}{2r_{0}^{2}}%
,\text{ }R_{\widehat{0}\widehat{2}\widehat{0}\widehat{2}}^{(s)}=0,R_{%
\widehat{0}\widehat{2}\widehat{0}\widehat{2}}^{(ex)}=\frac{\gamma -1}{%
2r_{0}^{2}}\left( \frac{v^{2}}{1-v^{2}}\right) .
\end{equation}%

We can see that, at the throat, the static curvature component $R_{\widehat{0}\widehat{2}\widehat{0}\widehat{2}}^{(s)}$ is $0$, while $R_{\widehat{0}\widehat{1}\widehat{0}\widehat{1}}$, as measured in the Lorentz-boosted frame, is negative for $0<\gamma <1$, which is consistent with the wormhole topology. However, the excess part $R_{\widehat{0}\widehat{2}\widehat{0}\widehat{2}}^{(ex)}$ can be very large if $v^{2}\rightarrow 1$. This largeness is a kinematic effect inherited from the Lorentz boost and not an enhancement of the  curvature.

\section{Influence of \boldmath$\gamma$ on the Fresnel Coefficients: the Tangherlini Formulation}\label{s4}
The approach of Tangherlini involves scales and clocks not affected by the medium; in addition, the  light falls normally on the  plane, semi-infinite, homogeneous, isotropic, and nonabsorbing real medium. When a photon is transmitted, the magnitude of its momentum $p^{\prime}$ inside the medium is related to the magnitude of its momentum $p$ in free space by the equation  \cite{Tangherlini:1975, Jones:1951, Jones:1954, Ashkin:1973}:
\begin{equation}
p^{\prime} = np,
\end{equation}%
where $n$ ($\geq$1) is the index of refraction. This equation is independent of the angle of incidence and is an experimental fact in an  ordinary optical medium. The de Broglie relation $p^{\prime} \lambda^{\prime} = p\lambda =$ constant (not Planck's constant) together with the reduction in the wavelength in the real optical medium, $\lambda ^{\prime} = \lambda /n$, which we  verify in the effective medium, lead to Equation (39); the unprimed quantities $p$ and $\lambda$ refer to those in free space (the absence of gravity or an effective medium).

Dynamically, the trajectory of a photon can be approximated by the following Hamiltonian, neglecting the dispersion (see \textit{Eq. (A1)} of Tangherlini  \cite{Tangherlini:1975}):%
\begin{equation*}
H^{\prime }=\frac{c_{0}}{n(\rho )}p^{\prime },
\end{equation*}%
where $n=n(\rho )$ is assumed to be a slowly varying refractive index for an ordinary optical medium oriented so that the boundary of the fluid is in the direction perpendicular to the photon rays.

Tangherlini's  \cite{Tangherlini:1975} statistical method introduces probability in the following novel way: the probability for a photon to be found either in a reflected or in a transmitted mode on the impinging surface has to be an {ensemble average}. The ensemble is meant to be a large number of identical media with the same refractive index. Associated with each incident photon, there is one ensemble. The condition $p^{\prime }\lambda ^{\prime }=p\lambda =$ constant does not involve Planck' s constant, and in this sense, the method introduces an indeterminacy of a pre-quantum type. The probability of reflection will be defined by the fraction of the total number of particles that are observed in the reflected mode and similarly for the probability of \mbox{the transmission.}

A static spherically symmetric wormhole in isotropic coordinates can be written as
\begin{equation}
d\tau ^{2} = \Omega ^{2}(\rho)dt^{2} - \Phi ^{-2}(\rho) [d\rho^{2} +
\rho^{2}\left( d\theta^{2} + \sin^{2}{\theta} d\varphi ^{2}\right)].
\end{equation}

Suppose $I$ is the flux of incident particles, $I_{R}$ is the flux of reflected particles, and $I_{T}$ is the flux of transmitted particles; then, number conservation implies
\begin{equation*}
I_{R}+I_{T}=I.
\end{equation*}%

In accordance with standard notation, the ratios $R=$ $I_{R}/I$ and $T=I_{T}/I$ are taken to define the probabilities of the reflection and transmission, respectively. These probabilities satisfy the conservation of the probability condition%
\begin{equation*}
R+T=1.
\end{equation*}%

Equating the average rate of energy delivered to the ensemble member in reflected and transmitted modes, Tangherlini  \cite{Tangherlini:1975} derives the coefficients:%
\begin{equation}
R=\frac{(n-1)^{2}}{(n+1)^{2}},\quad T=\frac{4n}{(n+1)^{2}},
\end{equation}%
which remain invariant under $n\rightarrow 1/n$.

The coefficients for the ingoing photon at the throat as observed by \textit{asymptotic observers (a.o.) }are given with $n$, using Equation (41):
\begin{eqnarray}
R_{\text{photon}}^{\text{a.o.}} &=&\frac{\left[ n(\rho _{\text{th}})-1\right]
^{2}}{\left[ n(\rho _{\text{th}})+1\right] ^{2}}, \\
T_{\text{photon}}^{\text{a.o.}} &=&\frac{4n(\rho _{\text{th}})}{\left[
n(\rho _{\text{th}})+1\right] ^{2}},
\end{eqnarray}%
where $n=\Omega ^{-1}\Phi ^{-1}$.

The coefficients for the ingoing photon pulse at the throat as observed by \textit{near-throat local observers (l.o.)} are given with $\widetilde{n}$ (=$n\Phi$), using Equation (41):
\begin{eqnarray}
\widetilde{R}_{\text{photon}}^{\text{l.o.}} &=&\frac{\left[ \widetilde{n}%
(\rho _{\text{th}})-1\right] ^{2}}{\left[ \widetilde{n}(\rho _{\text{th}})+1%
\right] ^{2}}, \\
\widetilde{T}_{\text{photon}}^{\text{l.o.}} &=&\frac{4\widetilde{n}(\rho _{%
\text{th}})}{\left[ \widetilde{n}(\rho _{\text{th}})+1\right] ^{2}}.
\end{eqnarray}

The idea is to find the coefficients of two pairs of equations, (42) and (43) and (44) and (45). Looking at these, we see that the formulation of the second pair is of a hybrid nature, as they have one leg in the effective medium defined by the index $n$ and the other in the metric function $\Phi$. For our purposes, it is enough to treat all of the equation pairs as merely some functions of $\rho$ and evaluate the coefficients for different observers~in different metrics. Thus, we convert solution (5) to isotropic form by the radial transform $r\rightarrow \rho$:
\begin{equation}
\rho =(\sqrt{r-r_{0}}+\sqrt{r-\gamma r_{0}})^{2}
\end{equation}%
and inverting, we find
\begin{equation*}
r=\frac{\rho ^{2}+2\rho r_{0}(1+\gamma )+r_{0}^{2}(1-\gamma )^{2}}{4\rho }.
\end{equation*}%

The thoat now occurs at
\begin{equation}
\rho _{\text{th}}=\left( 1-\gamma \right) r_{0}.
\end{equation}%

Then, metric (5) can be rewritten in the isotropic form
\begin{equation}
ds^{2}=\Omega _{\text{HKL}}^{2}dt^{2}-\Phi _{\text{HKL}}^{-2}(d\rho
^{2}+\rho ^{2}d\Omega ^{2}),
\end{equation}%
where
\begin{eqnarray}
\Omega _{\text{HKL}} &=&\text{exp}\left[ -\frac{4\rho r_{0}}{\rho ^{2}+2\rho
r_{0}(1+\gamma )+r_{0}^{2}(1-\gamma )^{2}}\right] , \\
\Phi _{\text{HKL}} &=&\frac{4\rho ^{2}}{\rho ^{2}+2\rho r_{0}(1+\gamma
)+r_{0}^{2}(1-\gamma )^{2}}.
\end{eqnarray}

Therefore, we find the refractive index, as perceived by asymptotic observers, to be
\begin{eqnarray}
n(\rho) &=& \frac{1}{\Omega _{\text{HKL}}\Phi _{\text{HKL}}} \\
&=&\frac{1}{4}\left[ 1+\frac{2r_{0}}{\rho }(1+\gamma )+\frac{%
r_{0}^{2}(1-\gamma )^{2}}{\rho ^{2}}\right] \text{exp}\left[ \frac{4\rho
r_{0}}{\rho ^{2}+2\rho r_{0}(1+\gamma )+r_{0}^{2}(1-\gamma )^{2}}\right],
\end{eqnarray}%
and similarly, the index as perceived by near-throat local observers is
\begin{eqnarray}
\tilde{n}(\rho ) &=& n\Phi _{\text{HKL}}  \\
&=&\text{exp}\left[ \frac{4\rho r_{0}}{\rho ^{2}+2\rho r_{0}(1+\gamma
)+r_{0}^{2}(1-\gamma )^{2}}\right] .
\end{eqnarray}%

At the throat, the indices (51)--(54) have the values
\begin{equation}
n(\rho _{\text{th}})=\frac{e}{1-\gamma },\quad \tilde{n}(\rho _{\text{th}})=e.
\end{equation}%

Thus, the reflection and transmission coefficients of the HKLWH for asymptotic observers are given by
\begin{equation}
R_{\text{photon}}^{\text{a.o.}}=\frac{\left( e-1+\gamma \right) ^{2}}{\left(
e+1-\gamma \right) ^{2}},\quad T_{\text{photon}}^{\text{a.o.}}=1-R_{\text{%
photon}}^{\text{a.o.}}
\end{equation}%
and for local observers, they are
\begin{equation}
\tilde{R}_{\text{photon}}^{\text{l.o.}}=\frac{\left( e-1\right) ^{2}}{\left(
e+1\right) ^{2}},\quad \tilde{T}_{\text{photon}}^{\text{l.o.}}=1-\tilde{R}_{%
\text{photon}}^{\text{l.o.}}.
\end{equation}%

From the above, it follows that
\begin{eqnarray}
\lim_{\gamma \rightarrow 1}{R_{\text{photon}}^{\text{a.o.}}} &=&1,\text{ }%
\lim_{\gamma \rightarrow 1}{T_{\text{photon}}^{\text{a.o.}}}=0 \\
\tilde{R}_{\text{photon}}^{\text{l.o.}} &=&0.213,\text{ \ }\tilde{T}_{\text{%
photon}}^{\text{l.o.}}=0.786.
\end{eqnarray}%

It is clear from Equation (57) that the Fresnel coefficients perceived by the local observers are independent of $\gamma$, whereas for the  asymptotic observers the Fresnel coefficients at the throat are the same as those of the Schwarzschild black hole horizon $R=1$, $T=0$, when $\gamma \rightarrow 1$. \textit{We take these values as base values representing with certainty a reflection characteristic of a stable horizon.} Thus, asymptotic observers are likely to identify the object as a black hole in the extreme limit $\gamma \rightarrow 1$, while local observers are more likely identify the object as an unstable wormhole, as the probability of the reflection $\tilde{R}_{\text{photon}}^{\text{l.o.}}$ is much less than the horizon value ${R_{\text{photon}}^{\text{a.o.}}}=1$. When $\gamma \rightarrow 0$, the Fresnel coefficients measured by the local and asymptotic observers, together with the corresponding observations, coincide.

\section{Conclusions}\label{s5}
The broad purpose of this paper was to study certain novel features of the HKLWH depending on the parameter $\gamma$. The behavior of different energy conditions for HKLWH were investigated in Section \ref{s2}, which showed that some were violated depending on the ranges of $\gamma$ and the size of the throat $r_{0}$. It was shown that the gravitational energy $E_{G}$ was negative on the positive side of the HKLWH meaning attractive gravity. Next, we studied the influence of $\gamma$ on the tidal forces in the Lorentz-boosted frame in Section \ref{s3}, which showed that the static curvature component $R_{\widehat{0}\widehat{2}\widehat{0}\widehat{2}}^{(s)}$ was $0$, while $R_{\widehat{0}\widehat{1}\widehat{0}\widehat{1}}$ as measured in the Lorentz-boosted frame was negative for $0<\gamma < 1$, which is consistent with the wormhole topology.

In Section \ref{s4}, we studied the influence of $\gamma$ on the Fresnel coefficients $R$ and $T$ by postulating that the identity of the observed object depended on the probabilistic outcome of the photon motion probing the object. The underlying idea was  explained in the introduction. The ``effective refractive medium'' with index $n(\rho)$ for a gravitational field could be valuable in the sense that it allows one to borrow wisdom from the phenomena occurring in an ordinary optical medium, for instance, the Fizeau effect  \cite{Nandi:2003}. The motion of light and particles in terms of refractive indices was previously carried out using a new formulation by Evans and Roenquist  \cite{Evans:1986, Rosenquist:1988, Evans:1990}. The application to gravity deriving exact general relativistic equations was conducted in  \cite{Evans:2001, Evans:1996a, Evans:1996b, Alsing:2001, Alsing:1998, Nandi:1995}. In the spirit of these works, we introduced the idea of a pre-quantum{ indeterminacy}, proposed by Tangherlini  \cite{Tangherlini:1975}, in identifying the gravitating source object to be either a BH or a WH. That means, we no longer asked what the object {truly} was but asked how the object was likely to be identified by differently located observers using the probe of the pre-quantum probability flow of photons. Thus, local observers may see the object as a HKLWH, while asymptotic observers are more likely to observe it as a BH horizon. Such a new possibility of different identifications of the same object, based on probabilistic outcomes of probes, does not seem to have been proposed heretofore.


\begin{thebibliography}{99}

\bibitem[Einstein(1935)]{Einstein:1935}
Einstein, A.; Rosen, N. The Particle Problem in the General Theory of Relativity. {\em Phys. Rev.} {\bf 1935}, {\em 48}, 73--77.

\bibitem[Ellis(1973)]{Ellis:1973}
Ellis, H.G. Ether flow through a drainhole: A particle model in general relativity. {\em J. Math. Phys.} {\bf 1973}, {\em 14}, 104. 

\bibitem[Bronnikov(1973)]{Bronnikov:1973}
Bronnikov, K.A. Scalar-tensor theory and scalar charge. {\em Acta Phys. Pol. B} {\bf 1973}, {\em 4}, 251--266.

\bibitem[Morris(1988)]{Morris:1988}
Morris, M.S.; Thorne, K.S. Wormholes in spacetime and their use for interstellar travel: A tool for teaching general relativity. {\em Am. J. Phys.} {\bf 1988}, {\em 56}, 395--412.

\bibitem[Barthiere(2018)]{Barthiere:2018}
Barthiere, C.; Sarkar, D.; Solodukhin, S.N. The fate of black hole horizons in semiclassical gravity. {\em Phys. Lett. B} {\bf 2018}, {\em 786}, 21--27.

\bibitem[Nandi(2004)]{Nandi:2004}
Nandi, K.K.; Zhang, Y.-Z.; Vijaya Kumar, K.B. Semiclassical and quantum field theoretic bounds for traversable Lorentzian stringy wormholes. {\em Phys. Rev. D} {\bf 2004}, {\em 70}, 064018.

\bibitem[Roman(1986)]{Roman:1986}
Roman, T.A. Quantum stress-energy tensors and the weak energy condition. {\em Phys. Rev. D} {\bf 1986}, {\em 33}, 3526--3533.

\bibitem[Karimov(2019)]{Karimov:2019}
Karimov, R.K.; Izmailov, R.N.; Nandi, K.K. Accretion disk around the rotating Damour–Solodukhin wormhole. {\em Eur. Phys. J. C} {\bf 2019}, {\em 79}, 952.

\bibitem[Harko(2009)]{Harko:2009}
Harko, T.; Kov\'{a}cs, Z.; Lobo, F.S.N. Can accretion disk properties distinguish gravastars from black holes? {\em Class. Quantum Grav.} {\bf 2009}, {\em 26}, 215006.

\bibitem[Tsukamoto(2017)]{Tsukamoto:2017}
Tsukamoto, N.; Gong, Y. Retrolensing by a charged black hole. {\em Phys. Rev. D} {\bf 2017}, {\em 95}, 064034.

\bibitem[Shaikh(2017)]{Shaikh:2017}
Shaikh, R.; Kar, S. Gravitational lensing by scalar-tensor wormholes and the energy conditions. {\em Phys. Rev. D} {\bf 2017}, {\em 96}, 044037.

\bibitem[Shaikh(2018)]{Shaikh:2018}
Shaikh, R. Shadows of rotating wormholes. {\em Phys. Rev. D} {\bf 2018}, {\em 98}, 024044.

\bibitem[Tsukamoto(2020)]{Tsukamoto:2020a}
Tsukamoto, N.; Kokubu, T. High energy particle collisions in static, spherically symmetric black-hole-like wormholes. {\em Phys. Rev. D} {\bf 2020}, {\em 101}, 044030.

\bibitem[Tsukamoto(2020)]{Tsukamoto:2020b}
Tsukamoto, N. Nonlogarithmic divergence of a deflection angle by a marginally unstable photon sphere of the Damour-Solodukhin wormhole in a strong deflection limit. {\em Phys. Rev. D} {\bf 2020}, {\em 101}, 104021.

\bibitem[Korolev(2020)]{Korolev:2020}
Korolev, R.; Lobo, F.S.N.; Sushkov, S.V. General constraints on Horndeski wormhole throats. {\em Phys. Rev. D} {\bf 2020}, {\em 101}, 124057.

\bibitem[Yusupova(2021)]{Yusupova:2021}
Yusupova, R.M.; Karimov, R.Kh.; Izmailov, R.N.; Nandi, K.K. Accretion flow onto Ellis–Bronnikov wormhole. {\em Universe} {\bf 2021}, {\em 7}, 177.

\bibitem[Stuchlik(2021)]{Stuchlik:2021}
Stuchl\'{i}k, Z.; Vrba, J. Epicyclic Oscillations around Simpson–Visser Regular Black Holes and Wormholes. {\em Universe} {\bf 2021}, {\em 7}, 279.

\bibitem[Paul(2021)]{Paul:2021}
Paul, B.C. Traversable wormholes in the galactic halo with MOND and non-linear equation of state. {\em Class. Quantum Grav.} {\bf 2021}, {\em 38}, 145022.

\bibitem[Radinschi(2021)]{Radinschi:2021}
Radinschi, I.; Grammenos, T.; Chakraborty, G.; Chattopadhyay, S.; Cazacu, M.M. Einstein and Møller Energy-Momentum Distributions for the Static Regular Simpson–Visser Space-Time. {\em Symmetry} {\bf 2021}, {\em 13}, 1622.

\bibitem[Alencar(2021)]{Alencar:2021}
Alencar, G.; Bezerra, V.B.; Muniz, C.R.; Vieira, H.S. Ellis–Bronnikov Wormholes in Asymptotically Safe Gravity. {\em Universe} {\bf 2021}, {\em 7}, 238.

\bibitem[Bronnikov(2021)]{Bronnikov:2021}
Bronnikov, K.A.; Kashargin, P.E.; Sushkov, S.V. Magnetized Dusty Black Holes and Wormholes. {\em Universe} {\bf 2021}, {\em 7}, 419.

\bibitem[Fabris(2021)]{Fabris:2021}
Fabris, J.C.; Gomes, T.A.O.; Rodrigues, D.C. Black Hole and Wormhole Solutions in Einstein–Maxwell Scalar Theory. {\em Universe} {\bf 2022}, {\em 8}, 151.

\bibitem[Jusufi(2022)]{Jusufi:2022}
Jusufi, K.; Kumar, S.; Azreg-Aïnou, M.; Jamil, M.; Wu, Q.; Bambi, C. Constraining wormhole geometries using the orbit of S2 star and the Event Horizon Telescope. {\em Eur. Phys. J. C} {\bf 2022}, {\em 82}, 633.

\bibitem[Sokoliuk(2022)]{Sokoliuk:2022}
Sokoliuk, O.; Praharaj, S.;  Baransky, A.; Sahoo, P.K. Accretion flows around exotic tidal wormholes---I. Ray-tracing. {\em Astronomy Astrophys.} {\bf 2022}, {\em 665}, A139. 

\bibitem[Tangherlini(1975)]{Tangherlini:1975}
Tangherlini, F.R. Particle approach to the Fresnel coefficients. {\em Phys. Rev. A} {\bf 1975}, {\em 12}, 139--147.

\bibitem[Bohm(1951)]{Bohm:1951}
Bohm, D. \textit{Quantum Theory}; Prentice-Hall: New York, NY, USA, 1951; p. 235.

\bibitem[Evans(2001)]{Evans:2001}
Evans, J.; Alsing, P.M.; Giorgetti, S.; Nandi, K.K. Matter waves in a gravitational field: An index of refraction for massive particles in general relativity. {\em Am. J. Phys.} {\bf 2001}, {\em 69}, 1103--1110.

\bibitem[Evans(1996)]{Evans:1996a}
Evans, J.; Nandi, K.K.; Islam, A. The optical–mechanical analogy in general relativity: New methods for the paths of light and of the planets. {\em Am. J. Phys.} {\bf 1996}, {\em 64}, 1404--1415.

\bibitem[Evans(1996)]{Evans:1996b}
Evans, J.; Nandi, K.K.; Islam, A. The optical-mechanical analogy in general relativity: Exact Newtonian forms for the equations of motion of particles and photons. {\em Gen. Relat. Gravit.} {\bf 1996}, {\em 28}, 413--439. 

\bibitem[Alsing(2001)]{Alsing:2001}
Alsing, P.M.; Evans, J.; Nandi, K.K. The Phase of a Quantum Mechanical Particle in Curved Spacetime. {\em Gen. Relat. Gravit.} {\bf 2001}, {\em 33}, 1459--1487. 

\bibitem[Alsing(1998)]{Alsing:1998}
Alsing, P.M. The optical-mechanical analogy for stationary metrics in general relativity. {\em Am. J. Phys.} {\bf 1998}, {\em 66}, 779--790. 

\bibitem[Evans(1986)]{Evans:1986}
Evans, J.; Rosenquist, M. ‘‘F=ma’’ optics. {\em Am. J. Phys.} {\bf 1986}, {\em 54}, 876--883. 

\bibitem[Nandi(2003)]{Nandi:2003}
Nandi, K.K.; Zhang, Y.-Z.; Alsing, P.M.; Evans, J.; Bhadra, A. Analogue of the Fizeau effect in an effective optical medium. {\em Phys. Rev. D} {\bf 2003}, {\em 67}, 025002. 

\bibitem[Nandi(2016)]{Nandi:2016}
Nandi, K.K.; Potapov, A.A.; Izmailov, R.N.; Tamang, A.; Evans, J.C. Stability and instability of Ellis and phantom wormholes: Are there ghosts? {\em Phys. Rev. D} {\bf 2016}, {\em 93}, 104044. 

\bibitem[Jones(1951)]{Jones:1951}
Jones, R.V. Radiation Pressure in a Refracting Medium. {\em Nature} {\bf 1951}, {\em 167}, 439--440. 

\bibitem[Jones(1954)]{Jones:1954}
Jones, R.V.; Richards, J.C.S. The pressure of radiation in a refracting medium. {\em Proc. R. Soc. A} {\bf 1954}, {\em 221}, 480--498. 

\bibitem[Ashkin(1973)]{Ashkin:1973}
Ashkin, A.; Dziedzic, J.M. Radiation Pressure on a Free Liquid Surface. {\em Phys. Rev. Lett.} {\bf 1973}, {\em 30}, 139--142. 

\bibitem[Harko(2008)]{Harko:2008}
Harko, T.; Kov\'acs, Z.; Lobo, F.S.N. Electromagnetic signatures of thin accretion disks in wormhole geometries. {\em Phys. Rev. D} {\bf 2008}, {\em 78}, 084005. 

\bibitem[Lobo(2017)]{Lobo:2017}
Lobo, F.S.N. \textit{Wormholes, Warp Drives and Energy Conditions}; Springer: Cham, Switzerland, 2017; pp. 11--254. 

\bibitem[Nandi(2004)]{Nandi:2004b}
Nandi, K.K.; Zhang, Y.-Z.; Vijaya Kumar, K.B. Volume integral theorem for exotic matter. {\em Phys. Rev. D} {\bf 2004}, {\em 70}, 127503. 

\bibitem[Harko(2013)]{Harko:2013}
Harko, T.; Lobo, F.S.N.; Mak, M.K.; Sushkov, S.V. Modified-gravity wormholes without exotic matter. {\em Phys. Rev. D} {\bf 2013}, {\em 87}, 067504. 

\bibitem[Parsaei(2019)]{Parsaei:2019}
Parsaei, F.; Rastgoo, S. Asymptotically flat wormhole solutions with variable equation-of-state parameter. {\em Phys. Rev. D} {\bf 2019}, {\em 99}, 104037. 

\bibitem[Rodrigues(2016)]{Rodrigues:2016}
Rodrigues, M.E.; Junior, E.L.B.; Marques, G.T.; Zanchin, V.T. Regular black holes in $f(R)$ gravity coupled to nonlinear electrodynamics. {\em Phys. Rev. D} {\bf 2016}, {\em 94}, 024062. 

\bibitem[Lynden-Bell(2007)]{Lynden-Bell:2007a}
Lynden-Bell, D.; Katz, J.; Bic\'{a}k, J. Energy and angular momentum densities of stationary gravitational fields. {\em Phys. Rev. D} {\bf 2007}, {\em 75}, 024040; Erratum in {\em Phys. Rev. D} {\bf 2007}, {\em 75}, E049901.  
{\em Phys. Rev. D} {\bf 2007}, {\em 75}, 024040. 

\bibitem[Nandi(2001b)]{Nandi:2001b}
Nandi, K.K.; Nayak, T.B.; Bhadra, A.; Alsing, P.M. Tidal forces in cold black hole spacetimes. {\em  Int. J. Mod. Phys. D} {\bf 2001}, {\em 10}, 529--538.

\bibitem[Rosenquist(1988)]{Rosenquist:1988}
Rosenquist, M.; Evans, J. The classical limit of quantum mechanics from Fermat’s principle and the de Broglie relation. {\em Am. J. Phys.} {\bf 1988}, {\em 56}, 881--882.

\bibitem[Evans(1990)]{Evans:1990}
Evans, J. Simple forms for equations of rays in gradient‐index lenses. {\em Am. J. Phys.} {\bf 1990}, {\em 58}, 773--778.

\bibitem[Nandi(1995)]{Nandi:1995}
Nandi, K.K.; Islam, A. On the optical–mechanical analogy in general relativity. {\em Am. J. Phys.} {\bf 1995}, {\em 63}, 251--256.

\end{thebibliography}
\end{document}